\documentclass[preprint]{aastex}
\shortauthors{CHUNG AND LEE}
\shorttitle{BINARY AND PLANETARY CENTRAL PERTURBATIONS}

\newcommand{\thetae}{\theta_{\rm E}}
\newcommand{\thetaeone}{\theta_{\rm {E,1}}}

\newcommand{\sbb}{s_{\rm b}}
\newcommand{\spp}{s_{\rm p}}
\newcommand{\qb}{q_{\rm b}}
\newcommand{\qp}{q_{\rm p}}
\newcommand{\Rb}{R_{\rm b}}
\newcommand{\Rp}{R_{\rm p}}
\newcommand{\sbhat}{{\hat s_{\rm b}}}
\newcommand{\delxib}{\Delta \xi_{\rm b}}
\newcommand{\deletab}{\Delta \eta_{\rm b}}
\newcommand{\delxip}{\Delta \xi_{\rm p}}
\newcommand{\deletap}{\Delta \eta_{\rm p}}

\begin{document}
\title{Distinguishing central perturbations by binary stellar and planetary systems under the moderately strong finite-source effect}

\author{Sun-Ju Chung and Chung-Uk Lee}
\affil{
Korea Astronomy and Space Science Institute, Hwaam-Dong,
Yuseong-Gu, Daejeon 305-348, Korea; sjchung@kasi.re.kr, leecu@kasi.re.kr
}


\begin{abstract}
We investigate high-magnification events caused by wide binary stellar and planetary systems under the moderately strong finite-source effect where the diameter of the source star is comparable with the caustics induced by a binary companion and a planet.
From this investigation, we find that a characteristic feature in the central perturbations induced by the binary systems commonly appears in a constant range where the size of the caustic induced by the binary companion is between 1.5 and 1.9 times of the diameter of the source, whereas in the central perturbations induced by the planetary systems the feature commonly appears in a range where the ratio of the size of the caustic induced by the planet to the source diameter changes with the planet/primary mass ratio.
High-magnification events caused by the binary and planetary systems with the characteristic feature produce a distinctive short-duration bump in the residuals from the single-lensing light curve, where the bump occurs near the time of peak magnification of the events.
Because of a well-known planet/binary degeneracy, we compare binary- and planetary-lensing events with the short-duration bump in the residuals.
As a result, we find the features of the binary-lensing events that are discriminated from the planetary-lensing events despite the moderately strong finite-source effect and thus can be used to immediately distinguish between the binary and planetary companions.
We also find the feature that appears only in binary-lensing events with a very low mass ratio or planetary-lensing events.
This implies that the lens systems with the feature have a very low mass binary companion (such as a brown dwarf) or a planet.

\end{abstract}


\section{INTRODUCTION}

The microlensing signal of a planet is a short-duration perturbation on the smooth standard light curve of the primary-induced lensing event that occurs on a background source star.
Microlensing experiments for the detection of a planet are carried out by two modes of observations.
One is survey mode (OGLE: Udalski 2003; MOA: Bond et al. 2002) which monitors a large area of sky and alerts ongoing events by analyzing data in real time.
The other is follow-up mode ($\mu$FUN: Dong et al. 2006, PLANET: Albrow et al. 2001, RoboNet: Burgdorf et al. 2007) which intensively monitors the alerted events.
Since the follow-up mode cannot monitor all the alerted events due to the limitation of the number of the telescopes, they tend to focus on high-magnification events for which the source star passes close to the primary star.
High magnification events are very sensitive for the detection of a planet because the central caustic induced by a planet is formed near the primary star and thus produce central perturbations in the lensing light curves \citep{griest98}.
Thus, 8 of 12 extrasolar planets found by microlensing have been discovered in the high-magnification events (Udalski et al. (2005); Gould et al (2006); Gaudi et al. (2008); Bennett et al. (2008); Dong et al. (2009); Janczak et al. (2010); Miyake et al.(2011)).

However, very close or very wide binary stellar systems can also produce central perturbations of high-magnification events.
Thus, the central perturbations induced by a very close or very wide binary companion and a planet are not generally distinguished without time consuming modeling.
Therefore, it is important to resolve the degeneracy between binary and planetary companions in high-magnification events (a well-known planet/binary degeneracy).
In addition, if the source star is much bigger than the central caustics induced by the binary companion and planet and thus the finite-source effect is strong, the central perturbations caused by the two caustics are greatly buried and the resulting light curves appear like that of a single lensing event induced by the primary star.
Thus, the planet/binary degeneracy under the strong finite-source effect becomes more severe.

Recently, high-magnification events with strong finite-source effects have been theoretically studied by \citet{han09a}, \citet{han09b}, and \citet{chung10}.
The results of these studies are that despite strong finite-source effects, some planetary-lensing events can be discriminated from binary-lensing events, and some events caused by triple lens systems composed of planets in wide binary stellar systems can be also discriminated from binary- and planetary-lensing events.
This is because the characteristic features of the planetary- and triple-lensing events with strong finite-source effects have been found.
However, since all studies of high-magnification events have been focused on cases where the source is much bigger or much smaller than the caustics induced by the binary companion and planet (cases of strong and weak finite-source effects), most of the high-magnification events are still difficult to be resolved without modeling.
Therefore, studies of the central perturbations of high-magnification events with other caustic/source size ratios are needed.
In this paper, we investigate high-magnification events caused by a binary companion and a planet under the moderately strong finite-source effect where the diameter of a source star is comparable with the caustics induced by the binary companion and the planet and compare the two central perturbation patterns.

The paper is organized as follows.
In Section 2, we briefly describe the properties of the central caustics caused by a binary companion and a planet.
In Section 3, we investigate the central perturbation patterns of binary and planetary lens systems under the moderately strong finite-source effect. We also compare the residual patterns of high-magnification events caused by the two lens systems.
We summarize the results and conclude in Section 4.

\section{CENTRAL CAUSTIC}

For a widely separated binary, the horizontal and vertical widths of the caustic caused by a binary companion are defined as the separations between the on- and off-axis cusps and are respectively expressed as
\begin{equation}
\delxib \simeq {4\gamma \over \sqrt{1 - \gamma}}, \quad \deletab \simeq {4\gamma \over \sqrt{1 + \gamma}}\ ;
\end{equation}
\begin{displaymath}
\gamma = {\qb \over{\sbhat^2}},\quad \qb  = {m_{2}\over{m_1}},
\end{displaymath}
where $\gamma$ is the shear induced by the binary companion, $\sbb$ is the projected primary-companion separation normalized by the Einstein radius corresponding to the total mass of th lens system, $\thetae$, and $m_1$ and $m_2$ are the masses of the primary and companion stars, respectively (Chang \& Refsdal 1979; Dominik 1999; Lee et al. 2008).
Here the $\rm ``hat"$ notation represents the length scale in units of the Einstein ring radius of the primary, $\thetaeone = \thetae [m_{1}/(m_{1}+m_{2})]^{1/2}$.
Thus, the vertical/horizontal width ratio, $\Rb$, is represented by
\begin{equation}
\Rb \simeq {\left( {1 - \gamma}\over{1 + \gamma}\right )^{1/2}}.
\end{equation}
Because $\sbb \gg 1$, $\gamma \ll 1$ and thus the width ratio becomes $\Rb \sim 1$.
As a result, the caustic for $\sbb \gg 1$ becomes a symmetric diamond-shaped caustic, while for $\sbb \rightarrow 1$ it becomes asymmetric.

On the other hand, the horizontal and vertical widths of the central caustic caused by a planet are respectively expressed by
\begin{equation}
\delxip \simeq {4\qp \over {(\spp - 1/\spp)^2}}, \quad
\end{equation}
\begin{displaymath}
\deletap \simeq {\delxip}{{(\spp - \spp^{-1})^{2}|\sin^{3}\phi|\over{( \spp + \spp^{-1} -2\cos{\phi})^{2}}}}\ ,
\end{displaymath}
where
\begin{displaymath}
\qp = {m_p\over{m_1} },
\end{displaymath}
\begin{displaymath}
\cos{\phi} = {{3\over 4}(\spp + \spp^{-1})\left\lbrace 1 - \sqrt{1 - {32\over 9}{1\over{(\spp + \spp^{-1})^{2}}}} \right\rbrace},
\end{displaymath}
where $\spp$ is the projected planet-primary separation normalized by $\thetae$ and $m_p$ is the planet mass \citep{chung05}.
Thus, the vertical/horizontal width ratio of the central caustic, $\Rp$, is represented by
\begin{equation}
\Rp \simeq  {(\spp - \spp^{-1})|\sin^{3}{\phi}|\over{(\spp + \spp^{-1} - 2\cos{\phi})^{2}}} \ .
\end{equation}
According to Equation (4), the shape of the central caustic depends only on the separation.
As a result, the width ratio of the central caustic is $\Rp < 1.0$, and the central caustic becomes more asymmetric as $\spp \rightarrow 1$.

\section{CENTRAL PERTURBATION PATTERN}

\subsection{\it Excess Map}

To investigate the central perturbation patterns of high-magnification events caused by wide binary stellar and planetary systems under the moderately strong finite-source effect, we construct magnification excess maps of the two systems.
The magnification excess is defined by
\begin{equation}
\epsilon = {A - A_{0} \over {A_0}}\ ,
\end{equation}
where $A$ and $A_0$ are the lensing magnifications with and without a companion, respectively.

Figures 1 and 2 show the magnification excess maps of wide binary stellar and planetary systems for various caustic/source size ratios and mass ratios.
In the two figures, the source radius changes for each column and the caustic size is fixed for each row.
In each map, the regions with blue and red colors represent the areas where the excess is negative and positive, respectively.
The color in the two figures changes into darker scales when the excess is $|\epsilon|$ = 1\%, 5\%, 15\%, 20\%, 22\%, 24\%, 26\%, 28\%, 30\%, and 32\%, respectively.
The dashed circle has a radius corresponding to that of the source and is located at the center of the caustic.
For limb darkening, we adopt a brightness profile for the source star of the form
\begin{equation}
{I(\theta)\over{I_0}} = 1 - \Gamma \left(1-{3\over{2}}{\rm cos}\theta \right ) - \Lambda \left(1-{5\over{4}}{\rm cos}^{1/2}\theta \right ) ,
\end{equation}
where $\Gamma$ and $\Lambda$ are the linear and square-root coefficients and $\theta$ is the angle between the normal to the surface of the source star and the line of sight.
We also adopt the coefficients of $\Gamma = -0.46$ and $\Lambda = 1.11$ (Han 2009; Chung \& Park 2010).

From the maps, we find that a characteristic feature in the central perturbations induced by the binary systems commonly appears in a constant range, $1.5 \lesssim \delxib/2\rho_\star \lesssim 1.9$, whereas in the central perturbations induced by the planetary systems the feature commonly appears in a range where the caustic/source size ratio changes with the planet/primary mass ratio.
For planetary systems with the mass ratio of $\qp = 5\times10^{-3}$, a characteristic feature appears in the range $1.8 \lesssim \delxip/2\rho_\star \lesssim 2.2$.
This feature is significantly formed at not only the region around caustic center (region $``1"$) but also other regions within the dashed circle (region $``2"$).
The regions $``1"$ and $``2"$ are marked in the maps with $\delxib/2\rho_\star = 1.7$ and $\delxip/2\rho_\star = 2.1$ of Figures 1 and 2.
The excesses of the two regions are smaller than those of the other regions within the circle.
This is because for the binary and planetary with the feature, the amount of the dominant negative excess within the area enclosed by the source is not much, but significantly, affected by the amount of the positive excess within the area when the center of the source is located within the circle.
For wide binary systems, the reason why the feature appears in a constant range, regardless of the mass ratio and separation of the binary, is that the shape of the caustic induced by a widely separated binary companion is symmetric ($\Rb \sim 1$).
On the other hand, for planetary systems, the reason why the feature appears in a range where the caustic/source size ratio changes with the planet/primary mass ratio, is that the planet-primary separation changes with the mass ratio for the same caustic/source size ratio and the shape of the caustic induced by a planet is asymmetric depending on the separation.
However, the characteristic feature regions become wider and the excesses of the regions become weaker ( $|\epsilon| \lesssim 5 \%$), as the source diameter becomes bigger than the caustics induced by the binary companion and planet.
As a result, the central regions of the binary and planetary systems start to show the features of binary and planetary systems with the strong finite-source effect (Han 2009; Han \& Kim 2009).
In addition, the strength of the perturbations outside the dashed circle is weaker than that of the perturbations affected by the weak finite-source effect, but the pattern of the perturbations is almost the same as that of the perturbations affected by the weak finite-source effect.
This means that events caused by the perturbations outside the dashed circle can be produced by events with the weak finite-source effect.

Figures 3 and 4 show the light curves and residuals from the single-lensing event resulting from the source trajectories presented in Figures 1 and 2.
Most of the perturbations induced by a binary companion and a planet in the lensing light curves are considerably washed out by the moderately strong finite-source effect, as shown in the figures.
However, events caused by the binary and planetary systems with the characteristic feature produce a distinctive signal near the time of peak magnification of the events.
The distinctive signal is the appearance of a short-duration bump in the residuals.
The duration of the bump is shorter or similar to those of two negative dips, where the negative dip occurs when the source enters and exits the caustic center.
For the binary and planetary systems with the feature, events where the source passes the caustic center or the center of the source passes the dashed circle generally produce the short-duration bump near the peak (see the middle panel of Figure 5).
The bottom panel of Figure 5 shows that events where the source does not pass the caustic center can be mimicked by events affected by the weak finite-source effect, as mentioned in the previous paragraph.
Thus, this study focuses on events passing the caustic center.
The duration of the bump increases as the source diameter increases, and thus two negative dips appear as two negative spikes and a flat region with the small excess of $|\epsilon| < 5\%$ appears between the two negative spikes.
As a result, the binary and planetary features that appear under the strong finite-source effect start to show in the residuals of the binary- and planetary-lensing events (Han 2009; Han \& Kim 2009; Chung \& Park 2010).

\subsection{\it Comparison with Binary- and Planetary-Lensing Residual Patterns}
\subsubsection{\it Symmetric Patterns}

The short-duration bump is a distinctive signal from events with weak and strong finite-source effects.
In this section, we thus investigate whether binary- and planetary-lensing events with the short-duration bump can be distinguished from each other.

Binary and planetary systems produce symmetric and asymmetric perturbations, respectively.
However, since the perturbations around the caustics induced by the binary companion and planet are still strong, binary- and planetary-lensing events with the moderately strong finite-source effect have generally asymmetric residual patterns.
As a result, most of binary and planetary events with the short-duration bump have asymmetric residual patterns, and thus the two kinds of events are generally difficult to distinguish.
Because of this fact, even though binary and planetary events with symmetric residual patterns are much rare than those with asymmetric residual patterns, they have higher probability of being resolved from each other than those with asymmetric patterns.
Therefore, we first compare binary and planetary events with symmetric residual patterns.

Figure 6 shows the magnification excess maps of binary and planetary systems together with the residuals resulting from the source trajectories presented in the individual maps. 
The mass ratios of the binary and planetary systems are $\qb = 0.5$ and $\qp = 5\times10^{-3}$, respectively.
The caustic/source size ratios of the binary systems I and II are $\delxib/2\rho_{\star} = 1.7$ and $1.88$, and the caustic/source size ratios of the planetary systems I and II are $\delxip/2\rho_{\star} = 2.1$ and $1.9$.
In Figure 6, we present all symmetric residual patterns that can occur in binary- and planetary-lensing events with the moderately strong finite-source effect.
However, while the residual patterns appear symmetric, they are not perfectly so.
There is a difference between the right and left excesses based on the time of $t = 0$, $|\Delta\epsilon_{RL}|$, but the difference is $|\Delta\epsilon_{RL}| < 5\%$ and it is very smaller than the whole depths of the residuals of $\gtrsim 40\%$.
Here the whole depth represents the difference between the positive and negative excesses with the maximum value in the residual.  
Thus, one defines that for binary- and planetary-lensing events with the moderately strong finite-source effect, the residual patterns with $|\Delta\epsilon_{RL}| < 5\%$ are symmetric.

For binary systems, there are six symmetric residual patterns, while for planetary systems there are two symmetric patterns.
This is because the central perturbations induced by a planet are symmetric with respect to only the planetary axis, and thus the source trajectories that have symmetric patterns are limited to the normal trajectories to the planetary axis.
Moreover, the normal trajectories passing the dashed semi-circle with $\hat{\xi} < 0.0$ do not produce the short-duration bump in the residuals.
This is because the perturbations of the center region (around the cusp at the planetary axis) of the semi-circle with $\hat{\xi} < 0.0$ are stronger than those of the other regions within the semi-circle, unlike the other semi-circle with $\hat{\xi} > 0.0$.
For example, trajectories IV, V, VI, and VII in the planetary maps I and II of Figure 6 are events passing the dashed semi-circle with $\hat{\xi} < 0.0$.
The four planetary-lensing events do not produce the short-duration bump near the peak, as shown in the panel for the trajectories.
Six symmetric patterns of the binary systems can classify into four patterns.
First, for the binary-lensing events of trajectories I and II, the short-duration bump and double positive spike simultaneously appear in the residual and the excesses before and after the positive spike are all negative.
Here the positive spike occurs at the moment just before the source enters into or just after the source exits from the caustic center.
The planetary-lensing event of trajectory I also produce the same residual feature with the above two binary events.
However, there is a prominent difference between the binary- and planetary-lensing events for trajectory I.
For the binary-lensing event, the depths of the negative excesses before and after the positive spike are deeper than those of two negative dips because the negative excesses outside the dashed circle are formed more strongly or similarly to the negative excess inside the circle.  
For the planetary-lensing event, the depths of the negative excesses before and after the positive excess are very shallower than those of the two negative dips.
This is because the source should always pass the region with a weak negative excess outside the circle in order to produce the symmetric residual feature with the short-duration bump (see trajectories I, II, and III in the planetary map I of Figure 6).
Thus, planetary-lensing events cannot produce the residual feature of the binary-lensing event for trajectory I.

Second, the residual feature of the binary-lensing event of trajectory III is the same as that of the planetary-lensing event of trajectory II.
The feature is that the short-duration bump and double positive spike appear together in the residual and the excesses before and after the positive spike are all positive.
In addition, a flat pattern of the positive excesses before and after the positive spike is also the same for the two events.
Thus, binary and planetary events with this feature are very difficult to resolve.
The residual feature of the binary-lensing event of trajectory IV, excluding the excess patterns before and after the positive spike, is the same as that of the binary event of the trajectory III.
In this event, the excess patterns before and after the positive spike show remarkable increasing and decreasing patterns, respectively.
The planetary-lensing event of trajectory III also produce the residual feature of this binary-lensing event.
Therefore, binary- and planetary-lensing events with the above two residual features are difficult to distinguish.

Third, the binary-lensing event of trajectory V has the short-duration bump and the positive excesses before and after the negative dip in the residual.
The positive spike does not appear in this event, since the source passes continuously the positive excess regions before and after crossing the caustic center.
This residual feature for planetary-lensing events cannot be produced.
This is because all source trajectories causing the short-duration bump always pass the positive excess regions only for the short duration before and after crossing the caustic center and thus produce the double positive spike in the residuals (see the panel for trajectories I, II, and III in the planetary map I of Figure 6).
Therefore, planetary events for which the short-duration bump occurs in the residual and the pattern of the residual is symmetric always have the double positive spike in the residuals. 

Fourth, the binary-lensing event of trajectory VI has almost the same residual feature as that of the binary event of the trajectory I.
However, unlike the event of the trajectory I, the event of the trajectory VI has the double positive-like spike in the residual.
The positive-like spike represents that the feature looks completely like the positive spike, but the excess of the spike is negative.
The positive-like spike occurs when the source passes the regions with small negative excess located around the fold caustics just outside the dashed circle.
Thus, the double positive-like spike occurs when the source passes diagonally two regions causing the positive-like spike, such as the binary event of the trajectory VI.
For the binary systems, four regions that cause the positive-like spike are formed around the fold caustics, as shown in the binary map II of Figure 6.
However, the four regions disappear as the caustic/source size ratio decreases (see Figure 1).
Planetary-lensing events can also produce the positive-like spikes in the residuals because two regions causing the positive-like spike are formed around the fold caustics toward the planet.
However, since one of two regions are not diagonally formed from the other one, as shown in the planetary maps I and II of Figure 6, the double positive-like spike cannot be produced in planetary-lensing events. 
Thus, planetary-lensing events cannot produce the residual feature of the binary-lensing event for trajectory VI.
As a result, we find three features of the binary-lensing events that are discriminated from the planetary-lensing events despite the moderately finite-source effect and thus can be used to immediately distinguish between the binary and planetary companions.
The three features are as follows.
\begin{enumerate}
\item
The short-duration bump and double positive spike appear together in the residual and the depths of the negative excesses before and after the positive spike are deeper than or similar to those of two negative dips.
\item
The short-duration bump appears in the residual and the excesses before and after the negative dip are all positive.
\item
The short-duration bump and double positive-like spike appear together in the residual.
\end{enumerate}
The patterns of the three binary-lensing features are all symmetric.

\subsubsection{\it Asymmetric Patterns}

As mentioned in Section 3.2.1, since most of binary- and planetary-lensing events with the short-duration bump are asymmetric, these two kinds of events are generally very difficult to distinguish.
However, because of different perturbation patterns between the binary and planetary systems with the moderately strong finite-source effect, it is possible that the two kinds of events despite the asymmetric pattern are discriminated from each other.
In this section, we thus choose a representative planetary-lensing event with an asymmetric pattern that is difficult to achieve in binary-lensing events and compare it with binary-lensing events.

The planetary-lensing event of trajectory I in the bottom panel of Figure 7 is the representative event with the asymmetric pattern.
For this planetary-lensing event, the short-duration bump and positive/positive-like spikes simultaneously appear in the residual and the excess before the positive spike is positive.
In addition, the depths of two negative dips are quite different ($|\Delta\epsilon_{RL}| \sim 10\%$), and thus the pattern of the residual feature of the planetary-lensing event is completely asymmetric.
On the other hand, in the case of binary-lensing events with the short-duration bump, the difference between the depths of two negative dips is mostly $|\Delta\epsilon_{RL}| < 5\%$.
This is because binary systems with the moderately strong finite-source effect produce symmetric perturbations inside the caustic.
For example, trajectories I, II, and III in three maps for $q = 0.1$ represent events that have the maximum difference between the depths of two negative dips in the individual maps.
All the maximum depth differences of the three events are $|\Delta\epsilon_{RL}| \lesssim 5\%$.
Since paths causing the maximum depth difference are quite limited, most of binary-lensing events with the short-duration bump have the two negative dips with $|\Delta\epsilon_{RL}| < 5\%$.
According to the definition of the symmetry mentioned in Section 3.2.1, the patterns of the two negative dips with $|\Delta\epsilon_{RL}| < 5\%$ are symmetric.
In addition, the binary-lensing events with the maximum depth difference between two negative dips do not produce simultaneously the positive-like spike and positive excess before or after the positive spike, such as the planetary-lensing event for trajectory I.
This is because paths causing the maximum depth difference do not correspond to paths causing both the positive-like spike and positive excess before or after the positive spike (see six trajectories in the top panel of Figure 7).
Among the events with both the positive-like spike and positive excess before or after the positive spike, trajectories IV, V, and VI in the maps for $q = 0.1$ represent events with the maximum depth difference.
All the maximum depth differences of the three events are $|\Delta\epsilon_{RL}| \lesssim 3\%$, as shown in the panel for the events.
Since the depth differences between two negative dips for the other trajectories are smaller than those of the above three trajectories, binary-lensing events with $ q = 0.1$ cannot produce the residual feature of the planetary-lensing event for trajectory I.
Binary-lensing events with $q > 0.1$ cannot also produce the planetary residual feature because the perturbations of the binary systems with $q > 0.1$ are more symmetric than those of the systems with $q = 0.1$.
Consequently, binary-lensing events with $q \geq 0.1$ cannot produce the completely asymmetric residual feature of the planetary-lensing event for the trajectory I.

However, since the shape of the caustic induced by a widely separated binary companion becomes asymmetric as the mass ratio and separation of the binary decrease, the central perturbations of binary systems with a very low mass ratio become asymmetric.
The second row from the top in Figure 7 shows that binary systems with very low mass ratios of $q = 0.04$ and $q = 0.03$ produce asymmetric central perturbations around the caustic like planetary systems.
Specifically, three maps for $q = 0.04$ show that the binary systems with $q = 0.04$ cannot still produce the completely asymmetric residual feature, which has not only the positive-like spike and the positive excess before or after the positive spike but also two negative dips with $|\Delta\epsilon_{RL}| > 5\%$ in the residual.
For example, trajectories VII, VIII, and IX in the three maps for $q = 0.04$ are representative events with the maximum depth difference between two negative dips among the events with both the positive-like spike and positive excess before or after the positive spike.
As shown in the panel for the three trajectories, the maximum depth differences are $|\Delta\epsilon_{RL}| < 5\%$ and thus the patterns of the negative dips are still symmetric. 
This means that binary-lensing events with $q = 0.04$ cannot still produce the complete asymmetric residual feature.
On the other hand, the panel for trajectory X in the map for $q = 0.03$ shows that the binary-lensing event of the trajectory X has the completely asymmetric residual feature.
However, considering the perturbations of this binary system, it seems that the binary event with the complete asymmetric residual feature for the binary systems with $q = 0.03$ can occur, but it is not easy to produce the feature.
This implies that the completely asymmetric residual feature appears well in binary systems with $q < 0.03$.
As a result, we find the residual feature that can appear only in binary systems with a very low mass ratio or planetary systems.
The feature is as follow.
\begin{enumerate}
\item[]
The short-duration bump and positive/positive-like spikes appear together in the residual and the excess before or after the positive spike is positive, and the difference between the depths of two negative dips is $|\Delta\epsilon_{RL}| > 5\%$.
\end{enumerate}
The pattern of this feature is completely asymmetric.
This implies that the lens systems with this feature have a very low mass binary companion (such as a brown dwarf) or a planet.

Considering that the photometric error of current follow-up observations reaches $\sim 1\%$ at the peaks of high-magnification events and the monitoring frequency of the observations is high, the short-duration bump and spike with deviation $\gtrsim 3\%$ can be readily detected.
However, we note that only a small fraction of binary- and planetary-lensing events with the moderately strong finite-source effect can be discriminated.
This is because the four features with symmetric and asymmetric patterns found by this study mostly occur when the center of the source passes around the caustic center.

\section{CONCLUSION}

We have investigated high-magnification events caused by wide binary stellar and planetary systems under the moderately strong finite-source effect.
From this study, we found that a characteristic feature in the central perturbations induced by the binary systems commonly appears in a constant range where the size of the caustic induced by a binary companion is between 1.5 and 1.9 times of the diameter of the source, whereas in the central perturbations induced by the planetary systems the feature commonly appears in a range where the ratio of the caustic induced by a planet to the source diameter changes with the planet/primary mass ratio.
High-magnification events caused by the binary and planetary systems with the characteristic feature produces a short-duration bump in the residuals from the single-lensing light curve, where the bump occurs near the time of peak magnification of the events.
Because of a well-known planet/binary degeneracy, we have compared the binary- and planetary-lensing events with the short-duration bump in the residuals.
From this, we found the features of the binary-lensing events that are discriminated from the planetary-lensing events despite the moderately strong finite-source effect and thus can be used to immediately distinguish between the binary and planetary companions.
We also found the feature that appears only in binary-lensing events with a very low mass ratio or planetary-lensing events.
This implies that the lens systems with the feature have a very low mass companion (such as a brown dwarf) or a planet.

\begin{figure}[t]
\epsscale{1.0}
\plotone{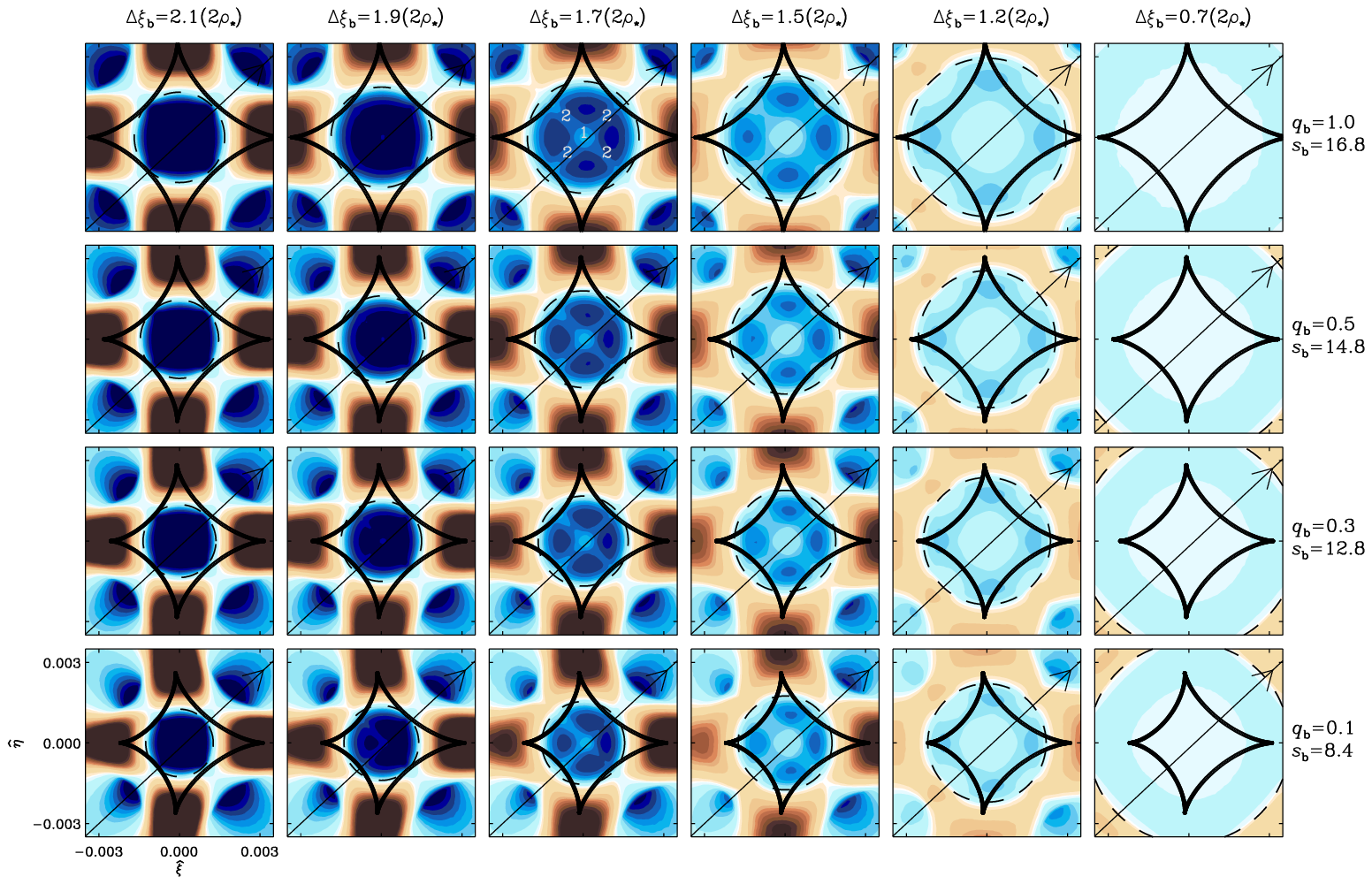}
\caption{\label{fig:one}
Magnification excess maps of wide binary stellar systems for various mass ratios and ratios of the caustic induced by a binary companion to the diameter of the source.
The coordinates ($\hat{\xi},\hat{\eta}$) represent the axes that are parallel with and normal to the binary axis and are centered at the effective position of the primary star.
Here, the notation with the hat represents the length scale normalized by the Einstein radius of the primary, $\thetaeone$.
In the map, the binary companion is located on the right, and $\qb$ and $\sbb$ are the mass ratio and separation of the binary, and $\delxib$ is the horizontal width of the caustic.
The color changes into darker scales when the excess is $|\epsilon|$ = 1\%, 5\%, 15\%, 20\%, 22\%, 24\%, 26\%, 28\%, 30\%, and 32\%, respectively.
The dashed circle represents the moment when the source is located at the center of the caustic.
The straight lines with arrows represent the source trajectories.
The regions marked as $``1"$ and $``2"$ represent the characteristic feature regions for the binary systems.
}
\end{figure}

\begin{figure}
\epsscale{1.0}
\plotone{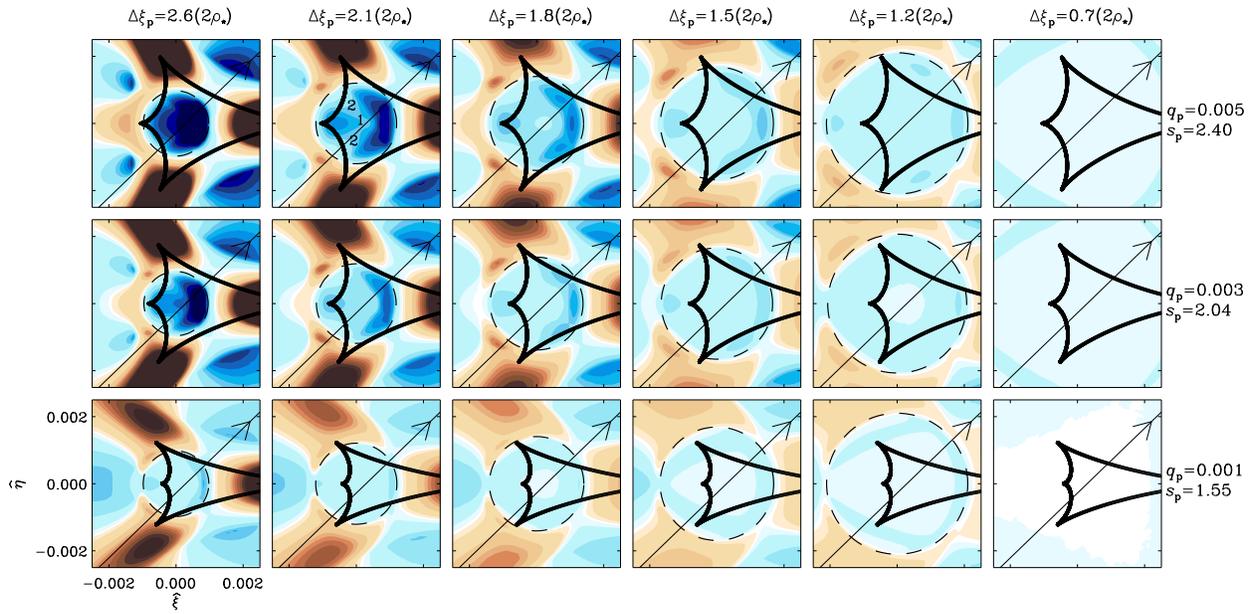}
\caption{\label{fig:two}
Magnification excess maps of planetary systems for various mass ratios and ratios of the caustic induced by a planet to the diameter of the source.
In the map, the planet is located on the right, and $\qp$ and $\spp$ are the planet/primary mass ratio and planet-primary separation, and $\delxip$ is the horizontal width of the caustic.
The regions marked as $``1"$ and $``2"$ represent the characteristic feature regions for the planetary systems.
}
\end{figure}

\begin{figure}[t]
\epsscale{1.0}
\plotone{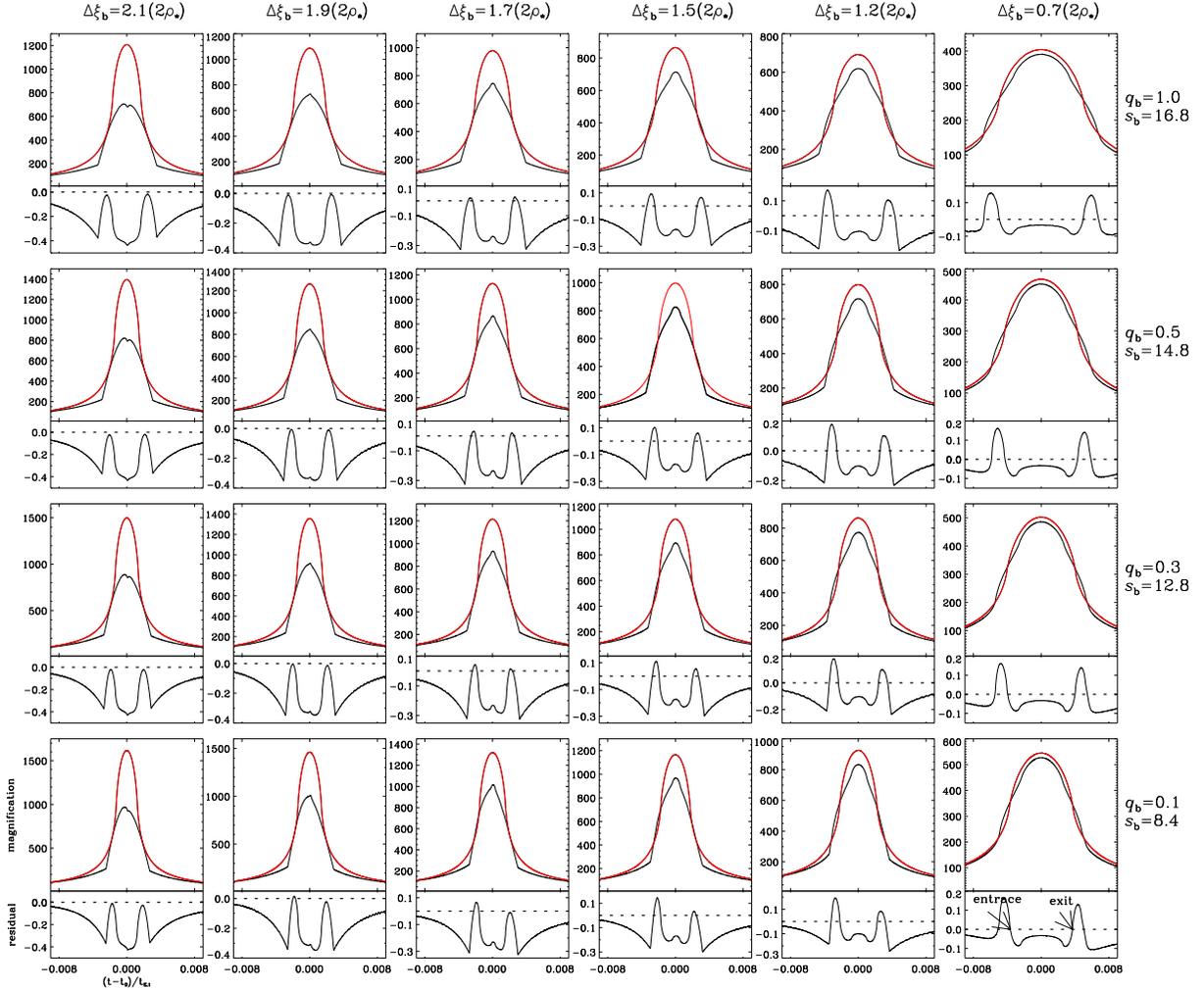}
\caption{\label{fig:three}
Light curves for the source trajectories presented in Figure 1.
In the upper panel, red and black curves represent the light curves of the binary- and single-lensing events, respectively.
The lower panel shows the residuals from the single-lensing light curve.
In the lower panel, the horizontal line indicates the magnification excess of $|\epsilon| = 0.0$.
The arrows in the right bottom panel represents the moments when the source enters and exits the dashed circle located at the caustic center in Figure 1.
}
\end{figure}

\begin{figure}[t]
\epsscale{1.0}
\plotone{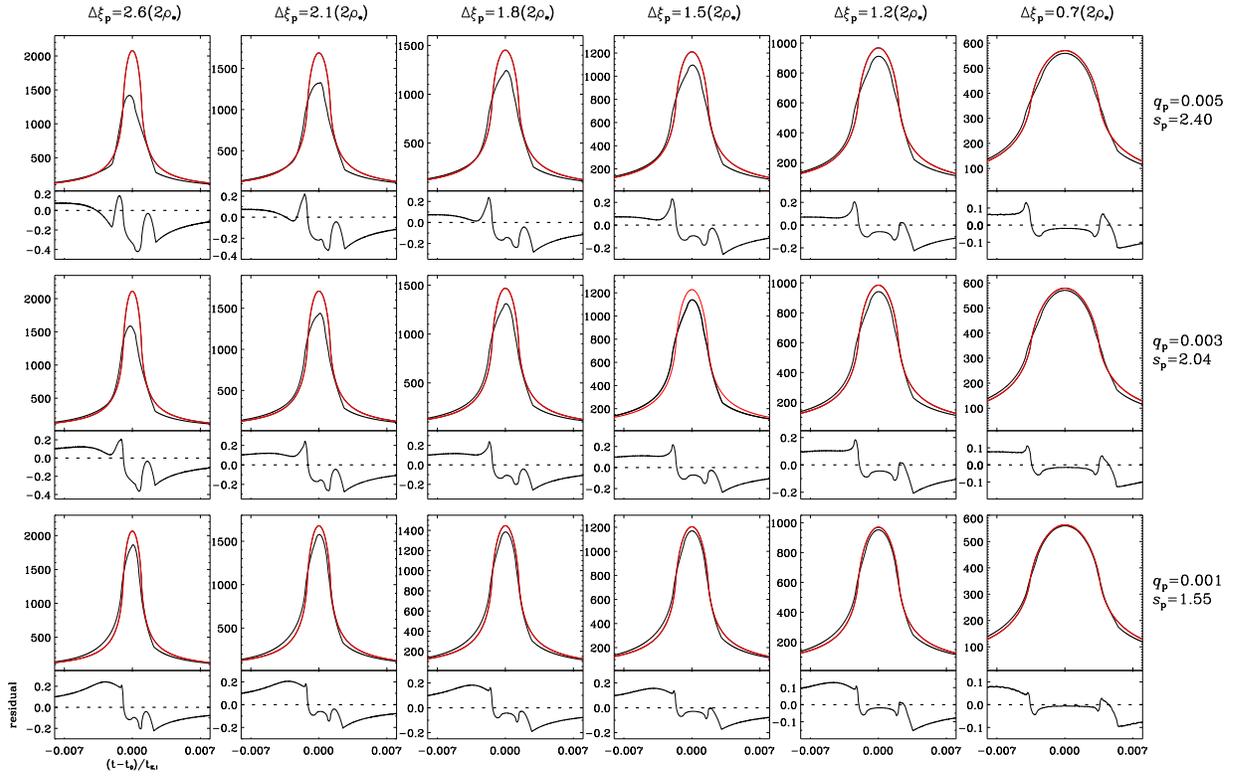}
\caption{\label{fig:four}
Light curves for the source trajectories presented in Figure 2.
In the upper panel, red and black curves represent the light curves of the planetary- and single-lensing events, respectively.
}
\end{figure}

\begin{figure}[t]
\epsscale{1.0}
\plotone{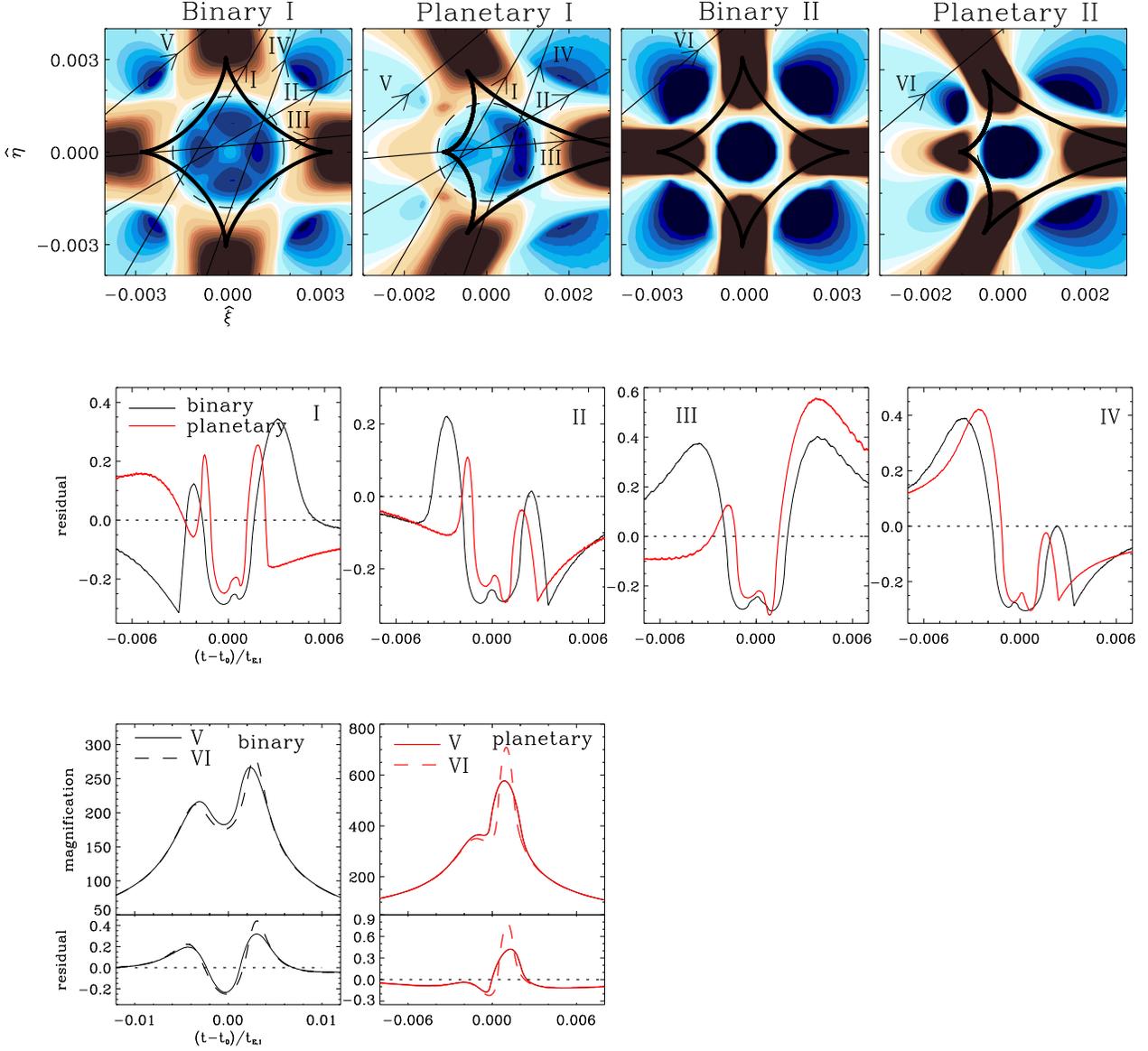}
\caption{\label{fig:five}
Magnification excess maps of binary and planetary systems with the moderately strong and weak finite-source effects and residuals resulting from the source trajectories presented in the individual maps.
The mass ratios of the binary and planetary systems are $\qb = 0.5$ and $\qp = 5\times10^{-3}$, respectively.
The caustic/source size ratios of the binary systems I and II are $\delxib/2\rho_{\star} = 1.7$ and $4.0$, and the caustic/source size ratios of the planetary systems I and II are $\delxip/2\rho_{\star} = 2.1$ and $4.0$.
}
\end{figure}

\begin{figure}[t]
\epsscale{1.0}
\plotone{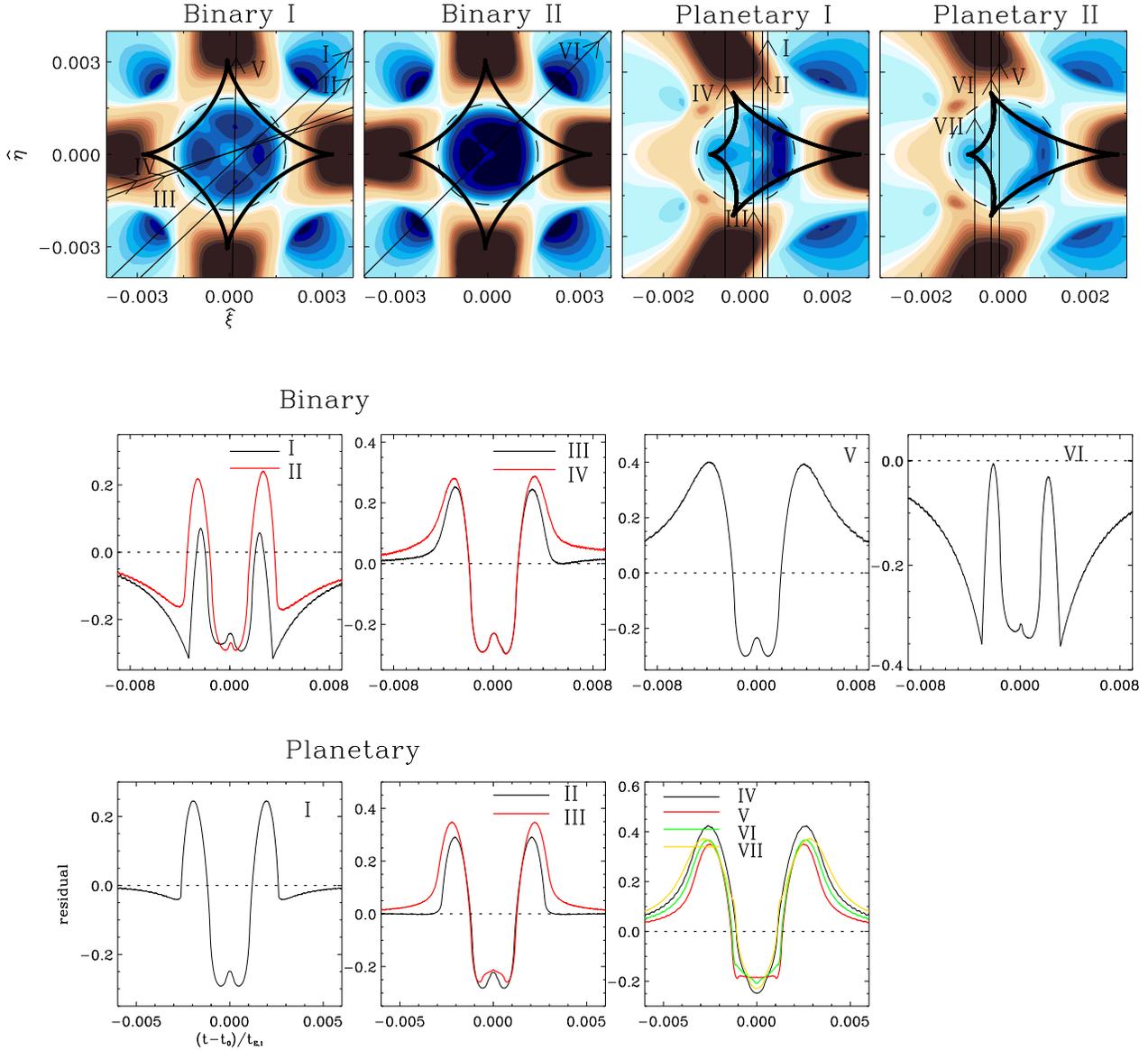}
\caption{\label{fig:six}
Magnification excess maps of binary and planetary systems together with the residuals resulting from the source trajectories presented in the individual maps.
The mass ratios of the binary and planetary systems are $\qb = 0.5$ and $\qp = 5\times10^{-3}$, respectively.
The caustic/source size ratios of the binary systems I and II are $\delxib/2\rho_{\star} = 1.7$ and $1.88$, and the caustic/source size ratios of the planetary systems I and II are $\delxip/2\rho_{\star} = 2.1$ and $1.9$.
}
\end{figure}

\begin{figure}[t]
\epsscale{1.0}
\plotone{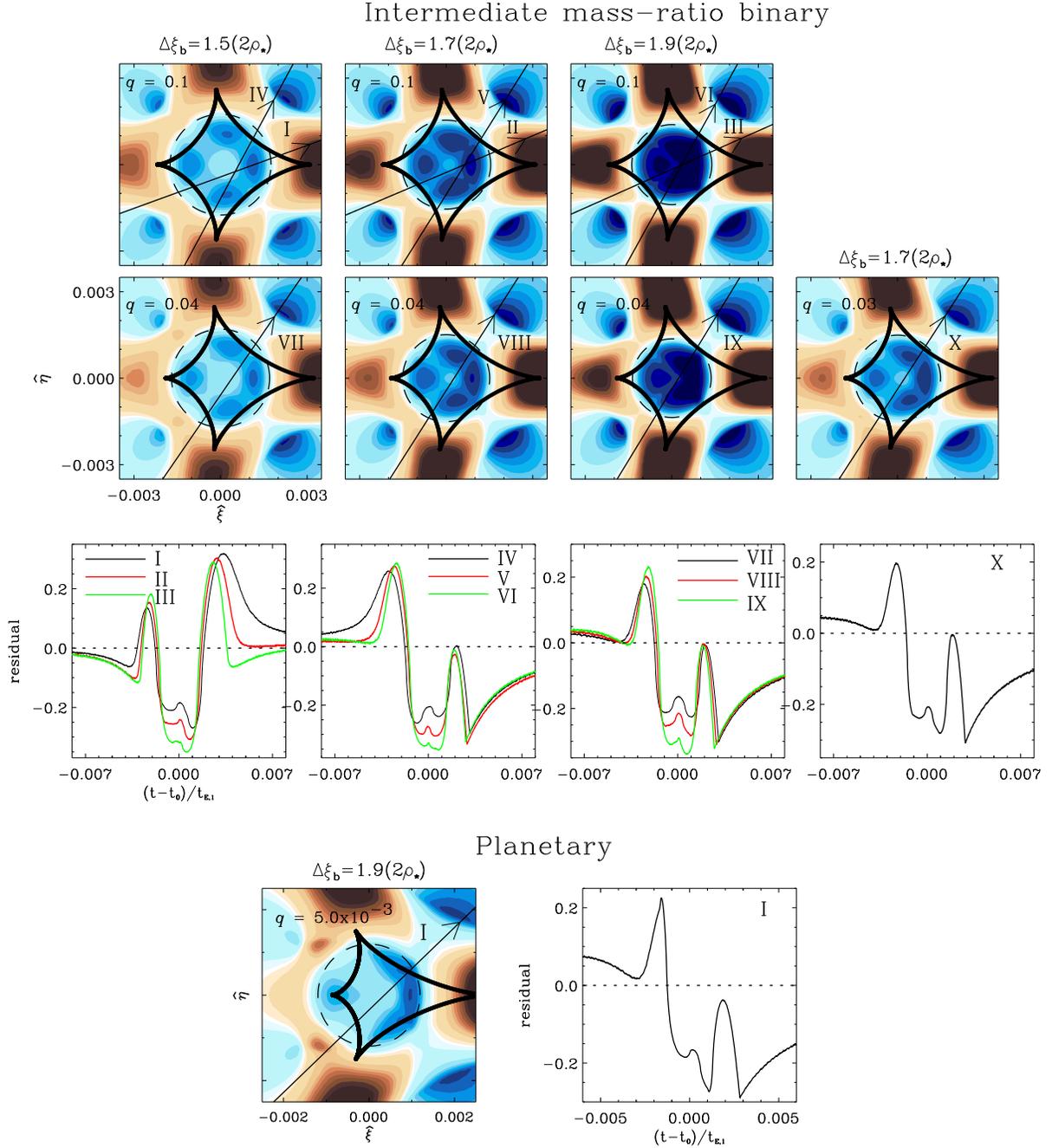}
\caption{\label{fig:seven}
Magnification excess maps of a planetary and various binary systems together with the residuals resulting from the source trajectories presented in the individual maps.
}
\end{figure}


We thank I.A. Bond for making helpful comments.


\begin{thebibliography}{99}
\frenchspacing

\bibitem[Albrow et al.(2001)]{albrow01}
Albrow, M.\ D., et al.\ 2001, \apj, 556, L113

\bibitem[Bennett et al.(2008)]{bennett08}
Bennett, D.\ P., et al.\ 2008, \apj, 684, 663

\bibitem[Burgdorf et al.(2007)]{burgdorf07}
Burgdorf, M.\ J., et al.\ 2007, \planss, 55, 582

\bibitem[Bond et al.(2002)]{bond02}
Bond, I.\ A., et al.\ 2002, \mnras, 331, L19

\bibitem[Chang \& Refsdal(1979)]{chang79}
Chang, K., \&  Refsdal, S.\ 1979, Nature, 282, 561

\bibitem[Chung et al.(2005)]{chung05}
Chung, S.-J., et al.\ 2005, \apj, 630, 535

\bibitem[Chung \& Park(2010)]{chung10}
Chung, S.-J., \& Park, B.-G.\ 2010, \apj, 713, 865

\bibitem[Dominik(1999)]{dominik99}
Dominik, M.\ 1999, A\&A, 349, 108

\bibitem[Dong et al.(2006)]{dong06}
Dong, S., et al.\ 2006, \apj, 642, 842

\bibitem[Dong et al.(2009)]{subo09}
Dong, S., et al.\ 2009, \apj, 698, 1826 

\bibitem[Gaudi et al.(2008)]{gaudi08}
Gaudi, B.\ S., et al.\ 2008, Science, 319, 927

\bibitem[Gould et al.(2006)]{gould06}
Gould, A., et al.\ 2006, \apj, 644, L37

\bibitem[Griest \& Safizadeh(1998)]{griest98}
Griest, K., \& Safizadeh, N.\ 1998, \apj, 500, 37

\bibitem[Han(2009)]{han09a}
Han, C.\ 2009, \apj, 691, L9

\bibitem[Han \& Kim(2009)]{han09b}
Han, C.,\& Kim, D.\ 2009, \apj, 693, 1835

\bibitem[Janczak et al.(2010)]{janczak10}
Janczak, J., et al.\ 2010, \apj, 711, 731

\bibitem[Lee et al.(2008)]{lee08}
Lee, D.-W.,\ Lee, C.-W.,\ Park, B.-G.,\ Chung, S.-J.,\ Kim, Y.-S.,\ Kim, H.-I., \& Han, C.\ 2008, \apj, 672, 623

\bibitem[Miyake et al.(2011)]{miyake11}
Miyake, N., et al.\ 2011, \apj, 728, 120

\bibitem[Udalski(2003)]{udalski03}
Udalski, A.\ 2003, Acta Astron., 53, 291

\bibitem[Udalski et al.(2005)]{udalski05}
Udalski, A. et al.\ 2005, \apj, 628, L109



\end{thebibliography}
\end{document}